# Strong terahertz radiation from relativistic laser interaction with solid density plasmas

Y. T. Li (李玉同)[1] [*], C. Li (李春)[1], M. L. Zhou (周木林)[1], W. M. Wang (王伟民)[1], F. Du (杜飞)[1], W. J. Ding (丁文君)[1], X. X. Lin (林晓宣)[1], F. Liu (刘峰)[1], Z. M. Sheng (盛政明)[1,2], L. M. Chen (陈黎明)[1], J. L. Ma (马景龙)[1], X. Lu (鲁欣)[1], Q. L. Dong (董全力)[1], Z. H. Wang (王兆华)[1], Z. Y. Wei (魏志义)[1], J. Zhang (张杰)[1,2] [*]

[1] *Beijing National Laboratory for Condensed Matter Physics, Institute of Physics, Chinese Academy of Sciences, Beijing 100190, China*
[2] *Shanghai Jiao Tong University, Shanghai 200240, China*

We report a plasma-based strong THz source generated by using intense femtosecond laser pulses to irradiate solid targets at relativistic intensity >$10^{18}$W/cm$^2$. Energies up to 50 μJ/sr per THz pulse is observed in the specular direction when the laser pulses are incident onto a copper foil at 67.5°. The source appears to be linearly polarized. The temporal, spectral properties of the THz are measured by a single shot, electro-optic sampling method with a chirped laser pulse. The THz radiation is attributed to the self-organized transient fast electron currents formed along the target surface. Such a strong THz source allows potential applications in THz nonlinear physics.

Terahertz (THz) radiation, which is located between the infrared light and microwaves in the frequency domain, attracts much interest due to increasingly wide applications [1]. Table-top THz sources can be generated by using optical pulses to pump various neutral materials such nonlinear crystals, silicon and so on [2,3,4,5,6,7]. With tilted laser wavefront scheme [8,9,10], up to 30 μJ THz pulse energy has been obtained. Plasma, which overcomes the damage problem of neutral materials when pump laser intensity is high, is a promising medium to generate strong THz radiation [11]. Observations of THz radiation from femtosecond laser-induced plasma filaments in air or other low density gases have been reported [12,13]. By applying a 'bias field', which is provided by an external electric field or an 'second harmonic optical field', such THz radiation can be much enhanced [14,15,16,17,18]. However, the radiation is subject to be saturated due to the limitation of laser intensity (typical <$10^{15}$W/cm$^2$, resulting from plasma defocusing effect [19]), low electron number involved, and THz re-absorption [18,20].

Compared with the neutral materials and plasma filaments in low density gases, almost arbitrarily high laser energies and intensities can be used in laser-solid interactions. A THz pulse with an energy of 1 μJ/sr was observed from solid targets irradiated by laser pulses at an intensity ~$10^{19}$ W/cm$^2$ [21]. Sagisaka *et al* also obtained a 0.5 μJ/sr THz pulse from thin foil targets [22]. The observations show the feasibility to produce strong THz sources with solid targets shined by high intensity laser pulses, though the generation mechanisms are still not well understood [23].

In this letter we present a ~50 μJ/sr strong THz pulse source generated in the relativistic laser-solid interactions. The polarization, temporal and spectral properties of the THz pulse are characterized. It is believed that a transient self-organized fast electron current along the target surface is mainly responsible for the generation.

The experiments were carried out using the Xtreme Light II (XL-II) Ti: sapphire laser system at the Institute of Physics, Chinese Academy of Sciences. A linearly-polarized laser pulse with an energy up to 150 mJ in 100 fs at 800 nm was focused onto a 30μm thick copper foil using an f/3.5 off-axis parabolic mirror. The target size was 100x200 mm. The diameter of the focal spot was ~5 μm. This gave a laser intensity up to 5×$10^{18}$ W/cm$^2$. The laser incidence angle was 67.5° with respect to target normal. An absolutely calibrated thermoelectric detector aligned in the laser specular direction was used to measure the THz pulse energy. The response range of the detector is from 0.3 to 21 THz. The collection solid angle was 0.11 sr. Silicon and plastic filters were set before the detector to transmit the THz radiation and block the visible light. The polarization of the THz pulse was measured by a wire-grid polarizer. The temporal waveform and the spectrum were measured by a modified, single-shot electro-optic method with a chirped laser pulse [24,25]. A 1mm thick ZnTe was used as the sampling crystal.

Figure 1 shows the measured waveform and normalized spectrum of the THz radiation when a *p*-polarized laser pulse is used. The duration of the THz pulse is about 1.5 ps. The peak frequency is about 0.5 THz. Note that there is a high frequency detection limit (5.3 THz) for the ZnTe crystal due to the first transverse optical phonon resonance [26]. Further studies are required to check if there are any THz components with frequency higher than the detection limit.

---

* To whom correspondence should be addressed. Email: ytli@aphy.iphy.ac.cn or jzhang1@sjtu.edu.cn.

Figure 2 shows the dependence of the THz pulse energy on pump laser energy. Each data point was taken by an average of ~10 shots. For the laser energy of 130 mJ the THz energy is 5.5 μJ in 0.11 sr, which corresponds to 50 μJ/sr. The energy of the THz radiation monotonously increases with laser energy. Unlike laser-gas interactions [18], no saturation effect is observed.

In the experiments, we also measured the angular distribution of THz radiation. The preliminary results show that THz is emitted into the whole $2\pi$ space in front of the target. For simplification, if we approximately take the THz radiation to be isotropic, the total conversion efficiency from laser energy to THz pulses is about $2.6 \times 10^{-3}$, which is similar to the two-color field [18] and titled wavefront [8] schemes. Higher THz pulse energies could be achieved if the laser energy or intensity is further increased.

Figure 3 shows the THz intensity as a function of the rotation angle of the wire-grid polarizer for *p*-polarized and *s*-polarized laser pulses, respectively. For both laser pulses, the polarization of the THz radiation is parallel to the laser incident plane. The ratio of the maximum to the minimum is about 10:1. The only difference between them is that the maximum THz intensity for the *s*-polarized laser pulses is lower by ~5 times than that for the *p*-polarized.

Two generation mechanisms have been proposed for the THz radiation in the laser-solid interactions, accelerated current arising from the longitudinal ponderomotive force [21] and "antenna mechanism" [22]. The two mechanisms cannot explain the features we have observed. The current driven by ponderomotive force will excite THz radiation in a radial polarization, not a linear polarization. According to "antenna" model, the THz frequency is determined by the target size. However, in our experiment we do not observe clear correlation between the THz frequency and the target size. This is also verified by Gao's experiment, in which a long wire is used as a target [23].

In the interaction of a high intensity relativistic laser pulse with a solid foil, energetic (fast) electrons with a temperature from several hundreds keV to several MeV can be generated [27,28].These fast electrons will form a transient current before transporting away from the interaction region in a time scale of few picoseconds. This current is probably responsible for the observed THz radiation. To show this, we have conducted simulations with a two-dimensional particle-in-cell code developed in our laboratory. The target is an overdense plasma slab located in the region 10-15 $\lambda_0$ in x-direction, 20-80 $\lambda_0$ in y-direction, where $\lambda_0$ is the laser wavelength (see Fig. 4). The target density is $6n_c$, where $n_c$ is the critical density. An 80fs p-polarized laser pulse is launched at the position (0,0) and irradiate on the front target surface at an incidence angle 60° with an intensity $2.6 \times 10^{18}$ W/cm$^2$. The size of the laser focal spot is 6 µm.

Figure 4 shows the simulated distributions of the electron current along the y-direction at the time of 50, 60, and 70 laser cycles. The time zero corresponds to the timing when the laser is launched. We can see that a well-confined electron current sheet is excited along the front target surface. The depth of the current sheet in x-direction is far less than 1$\lambda_0$. The current flows in y-direction with time. We find that the fast electron surface current is self-consistently induced due to spontaneous quasistatic magnetic and electrostatic fields produced at the target surface. This is confirmed in our previous experiments by observing the spatial distribution of fast electrons [29,30,31]. The surface current formation can be explained as follow. At the early stage of the interaction, when the fast electrons are accelerated into the target bulk, a strong MageGauss quasistatic magnetic field will be induced around the front surface. Simultaneously, a charge separation field is generated because some electrons are pulled out into vacuum by the laser electric field while ions do not move due to their big mass. Part of the fast electrons generated in the interaction will be reflected back to the vacuum by the quasistatic magnetic field. However, the negative electric field, whose peak position is slightly near vacuum relative to that of the magnetic field, will push them back to target again. The push-pull processes lead to a fast electron current along the surface. Therefore, a fast electron flow along the target surface is produced self-consistently. More details can be found in Refs. [29].

We believe that the surface current closely correlated with the observed THz radiation in our experiments. The dimension of the surface current in y-direction, which is similar to the interaction region 6μm/cos(60°), is much smaller than both the distance from the plasma to the detector, $R_0$, and $\sqrt{2R_0\lambda_T}$, where $\lambda_T$ is the wavelength of the THz radiation. Therefore, we can consider the current as a point source, i.e., ignore the phase difference of the radiation from different positions of the current.

Simulations and observations show that the surface current strength is increased with the laser intensity, $I_0$. Therefore, the relation between surface current and $I_0$ can be expressed as $\vec{J} = \alpha I_0 \exp(-t^2/\tau_0^2)\delta(\vec{r})\vec{e}_x$, where $1.665\tau_0$ is the full width at half maximum of the laser duration and $\alpha$ is a coefficient, which is dependent on the plasma and target parameters. According to the far field approximation, the THz electric field at the detector can be written as $\vec{E}_{THz} = \dfrac{2\alpha I_0 t \exp(-t^2/\tau_0^2)}{c^2 \tau_0^2 R_0 (1-\vec{n}\cdot\vec{\beta})} \vec{n}\times(\vec{n}\times\vec{e}_x)$, where $\vec{n}$ is the unit vector directing from the current center to the detector, $\vec{\beta}$ is the average velocity of the electrons in the current, and $c$ is the speed of light.

With the surface fast electron current source and the far field approximation, we can understand most features of the observed THz radiation. From the expression of $\vec{E}_{THz}$, we can see the THz amplitude increasing with the laser intensity, like the results in Fig. 2. The model indicates that the THz radiation in the laser incident plane is always $p$-polarization, no matter $p$- or $s$-polarized pump laser pulse is used. This is in agreement with the result in Fig. 3, in which the data were measured by the detector set in the incident plane. More laser energy is deposited into the plasma with a $p$-polarized laser pulse than with an $s$-polarized laser pulse [32]. This will lead to a stronger current and a consequent stronger THz radiation for the $p$-polarized laser pulse. In the approximation model the lifetime of the electron current is determined by the laser pulse duration. The current will disappear right after the laser is switched off. However, the relaxation of a real current in plasmas also depends on the plasma and target parameters. Typical lifetime of the current is in the time scale of ~ ps. Therefore, the electromagnetic radiation from such as a current falls in the THz regime.

In summary, a strong tabletop THz source from femtosecond laser-solid interactions has been generated at the relativistic laser intensity $>10^{18}$W/cm$^2$. The energy of a single THz pulse can be up to 50 μJ/sr with only ~100 mJ laser pulses. The total conversion efficiency from the laser pulse to the THz radiation in the whole space is >$10^{-3}$. A new self-organized fast electron current model is proposed to understand the generation. More powerful THz pulses with energies up to mJ or even higher are expected with the Petawatt laser systems available nowadays. Such sources would allow some new applications in the THz range, such as realtime imaging of dynamical process of materials with picosecond time resolution, excitation of nonlinear phenomena, etc.

## Acknowledgements

We thank S. C. Shi, C. L. Zhang for calibrating the detectors and filters, and L. Wang for discussions. This work is supported by the National Nature Science Foundation of China (Grant Nos. 10925421, 10734130), and National Basic Research Program of China (973 Program) (Grant No.2007CB815100 and 2007CB310406).

**Figure Captions**

Figure 1. Normalized spectrum of the THz radiation. The inset is the temporal waveform measured by a single-shot electric-optic method.

Figure 2. Absolute THz pulse energy as a function of the pump laser energy. The laser pulse is $p$-polarized. The THz pulse energy increases with the laser energy. 5.5 µJ THz radiation is obtained in a solid angle 0.11 sr for a 130 mJ pump laser pulse.

Figure 3. THz polarization. THz intensity as a function of the rotation angle of the polarizer for the $s$-polarized (blue circle, 0-180°) and $p$-polarized (red square, 180-360°) laser pulses. The data for the $s$-polarized laser pulses are artificially multiplied by a factor of 5 for clear comparison.

Figure 4. Electron current distributions at the target surface. An 80fs $p$-polarized laser pulse is incident from the left to right on the target surface at 60° with an intensity of $2.6 \times 10^{18}$ W/cm$^2$. The front target surface is located at x=10 $\lambda_0$. The electron current distributions at the time of 50 (a), 60 (b), and 70(c) laser cycles are shown.

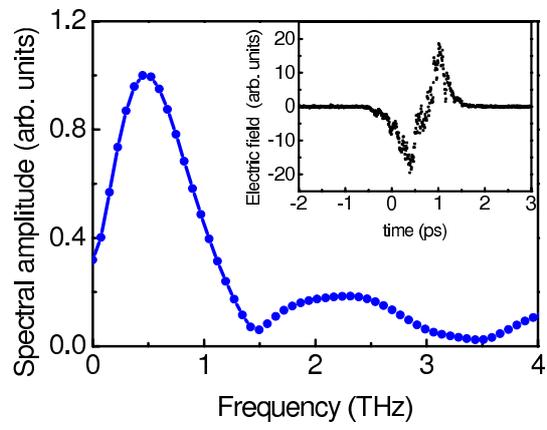

**Figure 1**

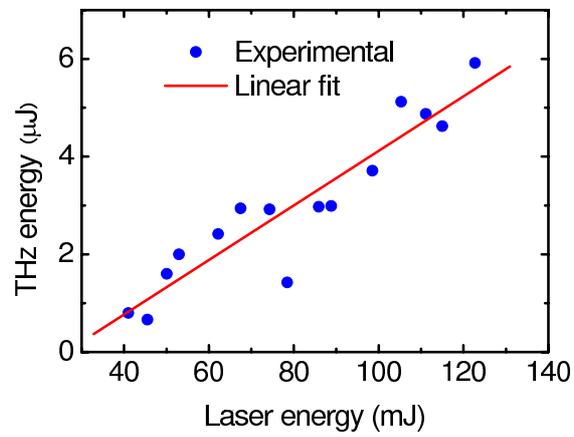

**Figure 2**

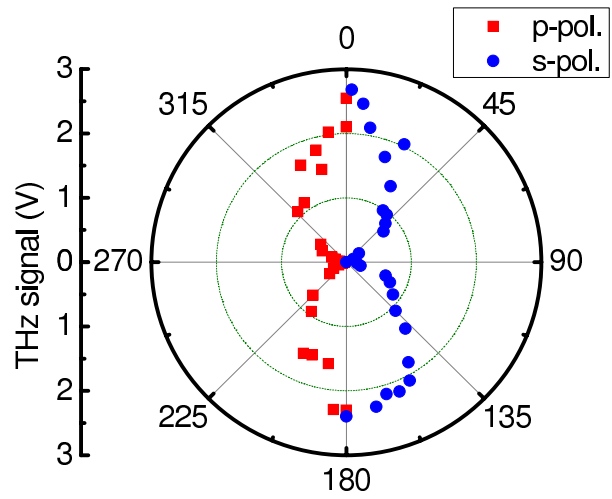

Figure 3

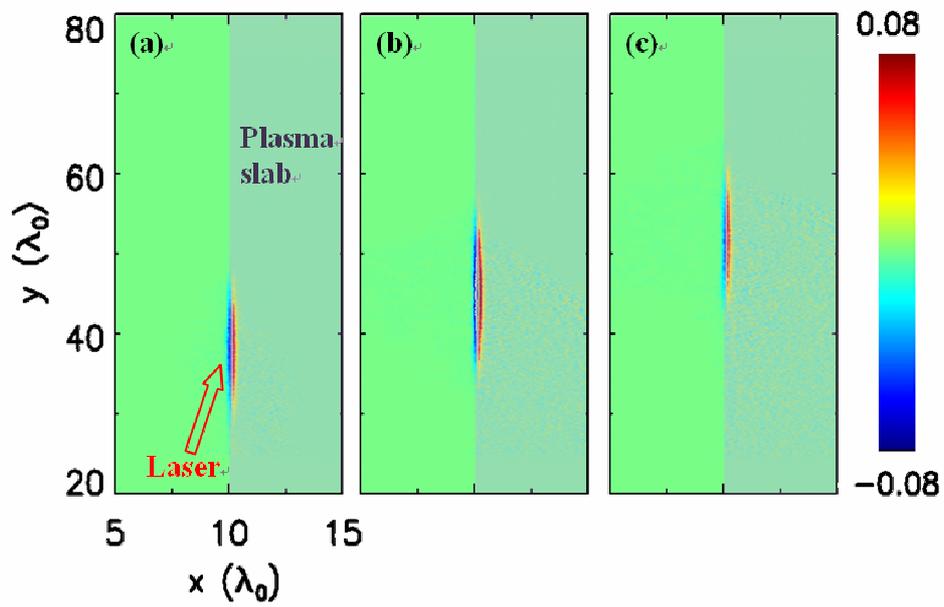

Figure 4